\documentclass{mn2e}
\usepackage{epsfig}
\usepackage{txfonts}
\newcommand{\HI}{H\,{\small I}}
\newcommand{\kms}{$\,$km$\,$s$^{-1}$}

\begin{document}

\title[FRB HI] 
{Neutral hydrogen absorption towards Fast Radio Bursts}

\author[Fender]
       {R. Fender$^1$\thanks{email: rob.fender@astro.ox.ac.uk} and T. Oosterloo$^{2,3}$ \\
       $^1$Astrophysics, Department of Physics, University of Oxford, Keble Road, Oxford OX1 3RH\\
       $^2$ASTRON, the Netherlands Institute for Radio Astronomy, Postbus 2, 7990 AA, Dwingeloo, The 
Netherlands.\\
$^3$Kapteyn Astronomical Institute, University of Groningen, P.O. Box 800,
9700 AV Groningen, The Netherlands}
\maketitle
\begin{abstract}
If Fast Radio Bursts (FRBs) are truly at astronomical, in particular cosmological, distances, they represent one of the most exciting discoveries in astrophysics of the past decade. However, the distance to FRBs has, to date, been estimated purely from their excess dispersion, and has not been corroborated by any independent means. In this paper we discuss the possibility of detecting neutral hydrogen absorption against FRBs both from spiral arms within our own galaxy, or from intervening extragalactic \HI\ clouds. 
In either case a firm lower limit on the distance to the FRB would be established. 
Absorption against galactic spiral arms may already be detectable for bright low-latitude bursts with existing facilities, and should certainly be so by the Square Kilometre Array (SKA).
Absorption against extragalactic \HI\ clouds, which would confirm the cosmological distances of FRBs, should also be detectable with the SKA, and maybe also Arecibo.
Quantitatively, we estimate that SKA1-MID should be able to detect extragalactic \HI\ absorption against a few percent of FRBs at a redshift $z \sim 1$. 
\end{abstract}

\section{Introduction}

One of the most exciting recent discoveries in astrophysics are the {\em Fast Radio Bursts}, apparently luminous coherent radio flashes at cosmological distances. These events were first reported in Lorimer et al. (2006), in data from the Parkes multibeam pulsar survey. To date, nine bursts from Parkes have been reported (Lorimer et al. 2006; Keane et al. 2009; Thornton et al. 2013; Petroff et al 2015; Ravi, Shannon \& Jameson 2015), as well as one from Arecibo (Spitler et al. 2014). These bursts are summarized in Table 1, see also summary in Keane \& Petroff (2014).
Their characteristics are of non-repeating, intrinsically narrow (intrinsic width $\leq 15$ms; perhaps as small as 1ms, see Keane \& Petroff 2014) radio bursts which have dispersion measures (DMs) which are greatly in excess of the expected maximum galactic DM along that line of sight. Under the assumption of a relatively low degree of dispersion per unit volume uniformly distributed in the intergalactic medium (IGM), their large excess DMs imply cosmological distances (up to redshift $z \sim 1$). 
At such distances the events have very high brightness temperatures ($\geq 10^{32}$ K) and may represent a new class of extreme astrophysical phenomena (e.g. Falcke \& Rezzolla 2014, Lyubarsky 2014, Mottez \& Zarka 2014, Cordes \& Wasserman 2015, Dolag et al. 2015). 
If truly at cosmologcal distances, their ability to probe the baryonic content and turbulence in the IGM, and maybe even the dark energy equation of state, makes them extremely exciting cosmological probes (see review in Macquart et al. 2015 and references therein), regardless of the underlying astrophysics of the event. 

However, their cosmological distances have not been, to date, corroborated in any other way. This is not surprising, given their poor localisation in single dish surveys which makes discovering a counterpart at higher frequencies very difficult (although there have been attempts: see Petroff et al. 2015, Ravi et al. 2015). For a while there was some doubt as to their astronomical reality due to their similarity to terrestrial/atmospheric events (`perytons', Burke-Spoalor et al. 2011, Katz 2014), which has not entirely gone away (Kulkarni et al. 2014). Loeb, Schvartzvald \& Maoz (2014) have suggested that the FRBs could originate from nearby flare stars, where the excess DM is local to the flare star, which would mean that FRBs were astrophysical, but were relatively local. Dennison (2014) argues against this. The local astrophysical scenario is also discussed in Kulkarni et al. (2014) and Burke-Spolaor \& Bannister (2014). Attempts to detect FRBs with an interferometer, which would provide a good localisation, have not yet been successful (Coenen et al. 2014, Law et al. 2015). 

The current situation is therefore that we have a population of radio transients, large numbers of which are likely to be detected by the Square Kilometre Array and its pathfinders/precursors (Hassall, Keane \& Fender 2013; Lorimer et al. 2013), and yet for which the distance, and hence the astrophysical origin and utility, remains uncertain. In this paper we demonstrate that the radio data themselves, by revealing neutral hydrogen absorption either from local spiral arms or absorbers at cosmological distances, have the potential to resolve this puzzle.

\begin{table}
\begin{tabular}{cccccc}
\hline
Burst & $l$ (deg) & $b$ (deg) & Est. z & \HI\ (MHz) & refs \\
\hline
FRB010125 &     356.6 & -20.0 & 0.6 & 890 & 1 \\
FRB010621 &     25.4 & -4.0 & 0.2 & 1200 & 2 \\
FRB010724 &     300.7 & -41.8 & 0.3 & 1100 & 3 \\
FRB110220 &     50.6  & -54.9 & 0.8 & 790 & 4\\
FRB110626 &     355.9 & -41.8 & 0.6 & 890 & 4\\
FRB110703 &     81.0  & -59.0 & 0.9 & 750 & 4\\
FRB120127 &     49.3  & -66.2 & 0.4 & 1000 & 4\\
FRB121102 &     175.0 & -0.2 & 0.3 & 1100 & 5 \\
FRB131104 &     260.6 & -21.9 & 0.6 & 890 & 6 \\
FRB140514 &     50.8 &   -54.6 & 0.4 & 1000 & 7 \\
\hline
\end{tabular}
\caption{
Summary of published FRBs, with galactic coordinates (typical positional error around half a degree), estimated redshift (large uncertainties), resulting estimated redshifted \HI\ frequency, and reference.
REFS: 1 = Burke-Spolaor \& Bannister (2014); 2 = Keane et al. (2012); 3 = Lorimer et al. (2007); 4 = Thornton et al. (2013); 5 = Spitler et al. (2014); 6 = Ravi, Shannon \& Jameson (2015); 7 = Petroff et al. (2015). Note that these the FRBs, and hence the references, are in chronological order {\em for the bursts}, not their discovery (the first burst discovered was FRB010724, by Lorimer et al. 2007). We caution that Bannister \& Madsen (2014) argue that FRB010621 is likely to be a galactic source. }
\end{table}

\section{\HI\ absorption against FRBs}

Were it possible to measure \HI\ absorption during the very short duration of FRB bursts, this could be a unique test of their distance, whether they are relatively nearby galactic phenomena or really at cosmological distances. 
How feasible is it to detect absorption features in the duration of a FRB? It turns out that it is in principle possible. In the following we assume that the DM to the FRB has already been accurately estimated and the data corrected for dispersion (`de-dispersed').

If the noise of the radio telescope, for a 1 msec integration at 1.420 GHz over a velocity range of 50 
\kms\ (corresponding to $\sim 240 \cdot (1+z)^{-1/2}$ kHz) , is $\sigma_\circ$ , the noise over a velocity interval $\Delta V$ \kms, as function of redshift $z$ of the absorber, is

\begin{equation}
\sigma = \sigma_\circ\ \big(\frac{50}{\Delta V}\big)^{1/2} t^{-1/2} (1+z)^{1/2} \ \  \mathrm{Jy}
\end{equation}

with $t$ the duration of the FRB in milliseconds. We have chosen 50 \kms\ because this is a representative width of \HI\ absorption detected in spectra of extragalactic sources (Prochaska et al. 2008) and because it broadly corresponds to the frequency resolution of the telescopes which have detected FRBs so far. 
For an FRB with a flux density $S$ this means that the noise level in optical depth for \HI\  absorption, with a width $\Delta V$, is

\begin{equation}
\sigma_\tau = \frac{\sigma_\circ}{S}\ \big(\frac{50}{\Delta V}\big)^{1/2} t^{-1/2} (1+z)^{1/2} \ \  \mathrm{Jy}.
\end{equation}

The column density of the absorbing cloud depends on the velocity-integrated optical depth and is given by $N_{\rm HI} [\mathrm{cm}^{-2}] = 1.8 \times 10^{18}(\int\tau dv)/T_{\rm spin}$, where $T_{\rm spin}$ is the spin temperature of the absorbing gas. The noise in the integrated optical depth is

\begin{equation}
\sigma_{\int\tau dv} = 50\ \frac{\sigma_\circ}{S}\ \big(\frac{50}{\Delta V}\big)^{1/2} W_{\rm 50}\ t^{-1/2} (1+z)^{1/2} \ \  \mathrm{km\ s}^{-1}.
\end{equation}

where $W_{50}$ is the width of the absorption profile in units of 50 \kms . So the 5-$\sigma$ detection limit $l$ for the integrated optical depth is

\begin{equation}
l = 250\ \frac{\sigma_\circ}{S}\ \big(\frac{50}{\Delta V}\big)^{1/2} W_{\rm 50}\ t^{-1/2} (1+z)^{1/2} \ \  \mathrm{km\ s}^{-1}.
\end{equation}

For existing telescopes such as the JVLA or the GBT, or the planned MeerKAT, the noise for a 1 msec integration is about 0.4-0.5 Jy over 50 \kms, depending somewhat on redshift. Assuming an FRB with a flux density of 10 Jy, we find, for $z = 0$, that the 1 msec detection limit in integrated optical depth for these telescopes is about 10-13 \kms, which is near the border of being interesting (see below). For the Parkes telescope, with which almost all known FRBs have been detected, the detection limit is around 15 \kms, suggesting that only in exceptional cases absorption will be detected. For the Arecibo telescope, the situation is better; the noise for a 1 msec observation is about 0.1 Jy, giving a detection limit of $\sim$2.5 \kms\ and it is not inconceivable that \HI\ absorption will be detected with this telescope, in particular against those FRBs at low Galactic latitude. For SKA1-Mid
\footnote{We use here the specification for SKA1-Mid as outlined in the SKA1 baseline design (2013): https://www.skatelescope.org/key-documents/}, the noise will be about 0.075 Jy implying a detection limit near 2 \kms. 
A main limitation of current telescopes is that their total bandwidth is relatively limited so that only a small redshift range is covered, reducing the likelihood to detect  absorption due to extragalactic \HI\ clouds. Given the observational setup used in the existing FRB detection, only Galactic absorption can be detected. With its wide observing band, SKA1, apart from its larger sensitivity, will be much more efficient in detecting \HI\ absorption.

\subsection{\HI\ absorption within the Milky Way}

\begin{figure*}
\includegraphics[scale=0.6]{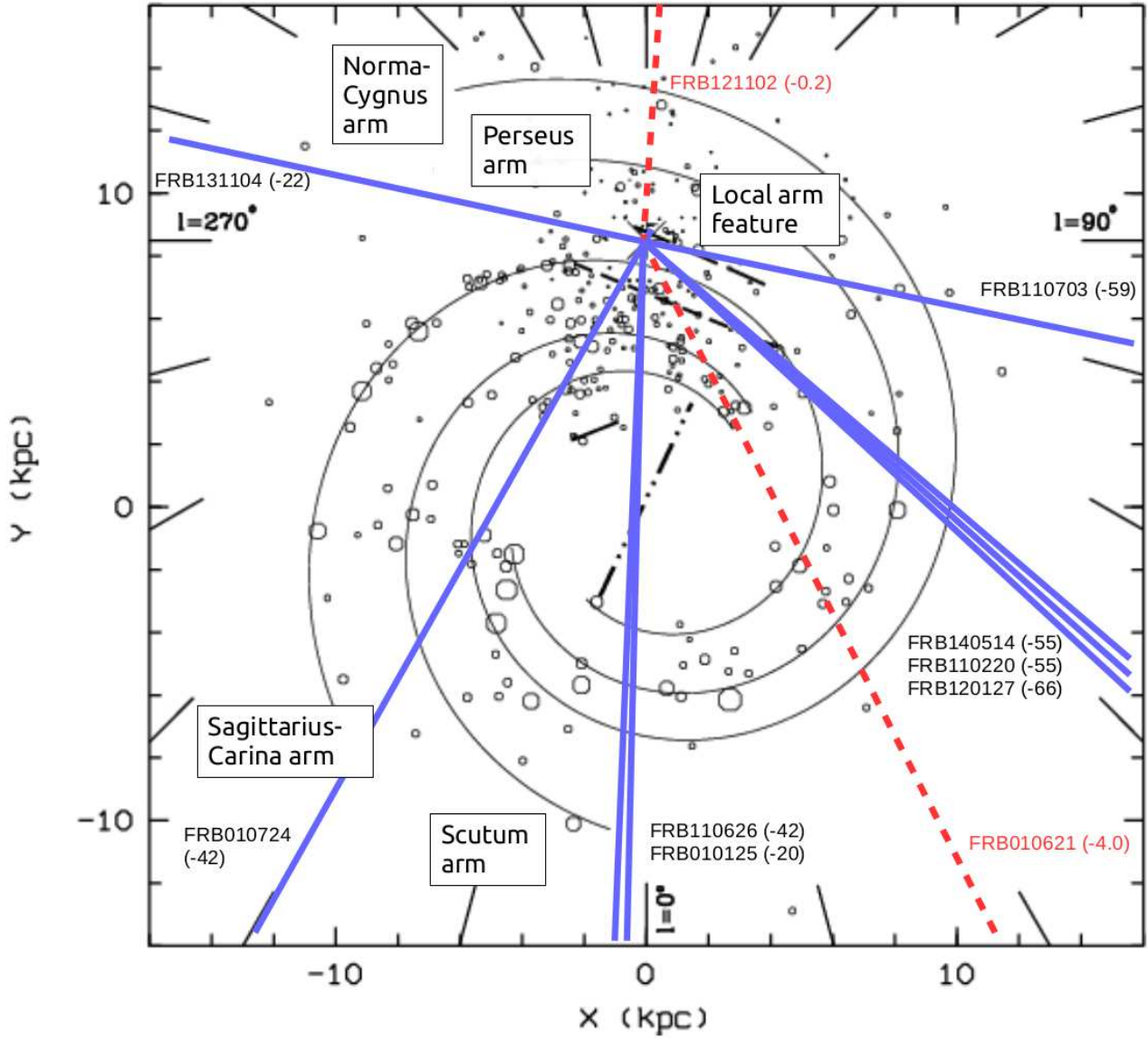}
\caption{Galactic longitiduinal directions (approximate) to FRBs overlaid on a schematic of the spiral arm structure inferred for the Milky Way (from Russeil 2003). Dashed (red) lines indicate low galactic latitudes ($|b| \leq 5^{\circ}$) which could potentially show some absorption against galactic spiral arms. The clustering of FRB detections in certain directions is almost certainly due to the biased observing directions of the Parkes surveys and has no astrophysical significance. The figures in parantheses after the FRB name are the galactic latitudes of the bursts.}
\label{spirals}
\end{figure*}

It may be possible to measure \HI\ absorption along the lines of sight to FRBs which are located in the galactic plane. The essence of the method is that radio spectra in a given direction towards a bright source will show \HI\ absorption at the velocities of the galactic spiral arms along the line of sight in that direction, and these features can be extracted by comparing the spectra of variables sources when they are bright and in quiescence.
This method has been used for several radio-bright X-ray binaries in the galactic plane (e.g. Braes et al. 1973, Goss \& Mebold 1977). The velocity shifts of the spiral arms are typically less than 100 km s$^{-1}$, meaning their signatures would lie within even the narrowest of bands which were centred on 1.4 GHz. In the case of FRB the data immediately before or after the burst should provide the `off' signal for comparison.

While this method does not give precise distance measurements, it does provide robust constraints, usually upper or lower limits, sometimes both. A bright FRB in the galactic plane, which had a spiral arm along the line of sight, and which did not show absorption against that spiral arm, would be likely constrained to be a nearby galactic object. This would be in direct contrast to the cosmological interpretation. Equally, an FRB which clearly showed absorption against a galactic spiral arm would be confirmed to be astrophysical and to lie at, at least, several kiloparsecs.

The \HI\ structure of the Milky Way is reviewed in e.g. Russeil (2003) and Kalberla \& Kerp (2009). In Fig \ref{spirals} we overplot the galactic longitudinal directions towards the FRBs listed in Table 1 on a schematic of the inferred spiral arm structure of the Milky Way. While most of the FRBs are well out of the plane, two of them, FRBs 010621 and 121102, have low enough galactic latitude that their lines of sight might have intersected neutral hydrogen in galactic spiral arms (see also Table 1). The line of sight to FRB010621 (Keane et al. 2012) intersects several galactic spiral arms which should have significant radial velocity components. However, the data are from Parkes which will not have a good signal to noise in the spectrum. FRB121102 was actually detected with Arecibo (Spitler et al. 2014) and intersects both the Norma-Cygnus and Perseus spiral arms in the galactic anticentre direction, although the radial velocity components are not going to be large. While it is currently a long shot, we recommend that all existing FRB data, perhaps especially these two, be checked for possible absorption features. 

Absorption against galactic spiral arms can produce strong features: the absorption spectrum towards Cir X-1 in Goss \& Mebold (1977) shows several features with integrated optical depths larger than 10 km s$^{-1}$. As outlined above, these should be easily detectable with the Square Kilometre Array, and even some of the larger existing (or under construction) radio telescopes.

\subsection{Extragalactic absorption}

If FRBs do lie at extragalactic distances, there is the possibility that \HI\ absorption  may be present in the spectrum of the FRB, as has been detected in the spectra of many extragalactic radio sources. This absorption can be either by the \HI\ of the ISM of the galaxy in which the FRB occurred (associated absorption, e.g., Morganti et al. 2001, Vermeulen et al. 2003, Curran \& Whiting 2010, Gereb et al. 2014), or by \HI\ clouds along the line of sight from the FRB toward the observer (intervening absorption, e.g., Kanekar et al. 2009, Curran et al.\ 2011). 

If the absorption can be established to be associated with the galaxy the FRB occurred in, it would immediately give the redshift at which the FRB occurred. Intervening absorption gives a lower limit to the redshift of the FRB. The probability of associated absorption occurs depends very strongly on the galaxy type in which the FRB is located. Although a significant fraction of early-type galaxies are now known to have significant amounts of neutral hydrogen (Serra et al. 2012), this \HI\ is often of low column density and  most of it is found at large radius where the stellar density is very low. The probability for detecting associated \HI\ absorption is therefore likely not to be very high. The situation  is very different for an FRB occurring in a spiral galaxy where it is quite possible the FRB is embedded in  \HI\ clouds of high column density.

To see which sensitivity is required and compare against the noise estimates made above, one can look at the statistics $dN/dX$, per unit comoving distance $X$ along any line of sight, of intervening \HI\ absorption systems exceeding a particular velocity integrated optical depth in radio spectra (Braun 2012). These statistics show that the number of absorbing systems with integrated optical depths larger than $\sim$10 \kms\ is rather low, but that for values below a few \kms, the probability increases. Therefore, in order to expect to detect a significant number of \HI\ absorptions, the detection limit $l$ of the radio telescope has to to be at most a few \kms. Interestingly, at low redshift, there are very few absorbing systems with  integrated optical depth below 1 \kms. This is mainly because at low \HI\ column densities, the effective spin temperature increases by about a factor 10 from a few hundred K to well above 1000 K which leads to a corresponding decrease in optical depth (Kanekar, Braun \& Roy 2011). Therefore a sensitivity that gives a detection limit below 1 \kms\ will not, at low redshift, lead to a larger number of detections of \HI\ absorption. For redshifts above about 1, this 'saturation' occurs around integrated optical depths of about 0.1 \kms. It therefore appears that with Arecibo or SKA1, one may realistically hope to detect \HI\ absorption in the spectrum of an extragalactic FRB. A major advantage of SKA1-Mid over current telescopes is that the instantaneous bandwidth is much larger so that a much larger redshift interval is covered, still with sufficient spectral resolution. The relatively narrow observing bands of Arecibo and Parkes, and the fact that these telescopes can only observe  at low redshifts, implies that the probability of detecting \HI\ absorption is much higher with observations with SKA1.

\begin{figure}
\includegraphics[width=\hsize]{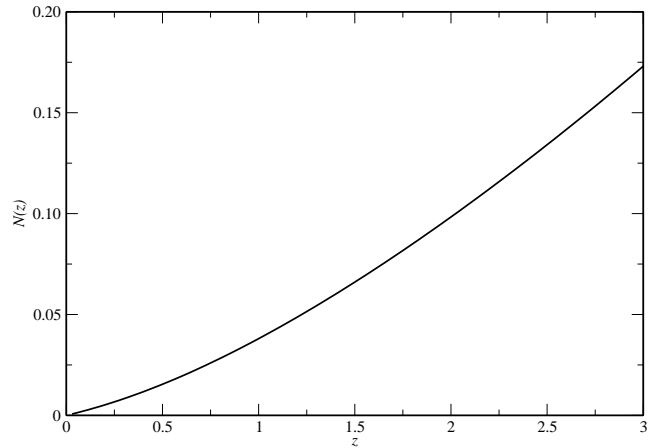}
\caption{Expected number of \HI\ absorbing system with integrated optical depth $\int \tau dv > 2$ \kms\, the estimated sensitivity of SKA1-MID, in the spectrum of an FRB at redshift $z$.}
\label{Nz}
\end{figure}

To get an idea of the number of expected intervening absorptions, we note that, interestingly, for \HI\ absorbing systems with integrated optical depths
around 2 \kms, the probability to detect absorption at this, or
higher, levels, is independent of redshift and is about 0.015 per unit
comoving distance. Assuming $dN/dX$ is independent of redshift, and using $dX = \frac{H_\circ}{H(z)} (1 + z)^2\, dz$, it is straightforward to compute the probability that intervening absorption occurs, as function of redshift of the FRB. This is plotted in Fig.\ \ref{Nz}. This figure shows that for a detection limit near 1 \kms, which will be achievable with SKA1, there is a significant probability to detect \HI\ absorption if FRBs are at cosmological distances with observations using the the lowest frequency band of SKA1 (`Band 1', which covers the redshift interval $0.35 < z < 3.0$ in a single observation). Further into the future, the improved sensitivity of the full SKA, which will be about a factor 10 better than that of SKA1, the number of detected intervening \HI\ absorptions will not increase at low redshift, but for redshifts higher than about 1, the probability will be significantly higher compared to SKA1.

\section{Discussion}

If FRBs really are at cosmological distances, then they would represent one of the most exciting discoveries in astrophysics of the past decade. However, this has yet to be definitively established. We have shown that it may be possible, with new high sensitivity radio telescopes, to detect \HI\ absorption against a subset of these bursts. The data would have to be de-dispersed and flux calibrated, but this is already feasible at some level (e.g. Lorimer et al. 2006, Spitler et al. 2014, Ravi et al. 2015, all report spectral indices for FRBs across a relatively narrow band). The strongest \HI\ absorption signals could potentially come from absorption against spiral arms in our own galaxy, although this is hampered by the dearth of FRBs at low galactic latitudes (Petroff et al. 2014).

If we do detect \HI\ absorption, then we constrain the distance to FRBs, even for bursts with total durations as short as milliseconds. If absorption is detected corresponding to \HI\ at cosmological distances, it would provide a lower limit to the redshift of the FRB. To establish whether the \HI\ redshift corresponds to the redshift of the FRB (associated absorption), follow up observations would be required to see whether galaxies are found at or near the location of the FRB that have the same redshift as the \HI\ absorption. If no extragalactic \HI\ absorption is detected for a sample of FRBs (of order 100), this would suggest they are not, after all, at cosmological distances. We note that since of order one FRB {\em per day} is expected to be detected with the first phase of the Square Kilometre Array (specifically SKA1-MID, see Hassall, Keane \& Fender 2013; Lorimer et al. 2013; Macquart et al. 2015), a clear result should be available within a year of observations.
Of course it is still to be hoped that multiwavelength counterparts to FRBs can be detected which will allow further independent distance estimates, but the technique outlined within this paper should provide a firm test of the astronomical distances of FRBs. In the meantime, it would be prudent to check all existing FRB spectra for any evidence of absorption.

\section*{Acknowledgements}

RF would like to acknowledge useful conversations with Evan Keane and Naomi McClure-Griffiths. RF was partly funded by ERC Advanced Investigator Grant 267607 `4 PI SKY'.

\end{document}